\documentclass[twocolumn]{aastex63}

\usepackage{import}
\usepackage{color}
\usepackage{listings}
\usepackage{amsmath}
\usepackage{graphicx}
\usepackage[normalem]{ulem}

\begin{document}

\shortauthors{Covarrubias, Wang \& Blunt}
\title{N-body Interactions will be Detectable in the HR-8799 System within 5 years with VLTI-GRAVITY}

\author{Sofia Covarrubias}
\affiliation{Department of Physics and Astronomy, California State University, Los Angeles}
\affiliation{Department of Astronomy, California Institute of Technology, Pasadena, CA, USA}

\author{Sarah Blunt}
\affiliation{Department of Astronomy, California Institute of Technology, Pasadena, CA, USA}
\affiliation{NSF Graduate Research Fellow}
\affiliation{Denotes Equal Coauthorship}

\author{Jason J. Wang}
\affiliation{Department of Astronomy, California Institute of Technology, Pasadena, CA, USA}
\affiliation{51 Pegasi b Fellow}
\affiliation{Denotes Equal Coauthorship}

\begin{abstract}
While Keplerian orbits account for the majority of the astrometric motion of directly-imaged planets, perturbations due to N-body interactions allow us to directly constrain exoplanet masses in multiplanet systems. This has the potential to improve our understanding of massive directly-imaged planets, which nearly all currently have only model-dependent masses. The VLTI-GRAVITY instrument has demonstrated that interferometry can achieve 100x better astrometric precision \citep{Grav_coll_2019} than existing methods, a level of precision that makes detection of planet-planet interactions possible. In this study, we show that in the HR-8799 system, planet-planet deviations from currently used Keplerian approximations \citep{2021Lacour} are expected to be up to one-quarter of a milliarcsecond within five years, which will make them detectable with VLTI-GRAVITY. Modeling of this system to directly constrain exoplanet masses will be crucial in order to make precise predictions.

\end{abstract}

\section{Introduction}
\label{sec:intro}

Orbits are essential to understanding the formation and evolution of exoplanet systems. In particular, orbital elements such as the inclination angle and eccentricity directly reflect the dynamical history of the system \citep{2020AJ}. Recently, planet-planet interactions have been used to measure the dynamical masses of directly-imaged exoplanets \citep{2021Lacour}, which has the potential to address one of the major limitations of using direct imaging: that planet masses are currently model-dependent. This new method can be used to compare measured masses against those predicted by formation models and even discover new planets using existing data. In the era of extremely precise astrometry, using planet-planet interactions to measure dynamical masses is not only possible, but may be a more effective method than using radial velocities and/or absolute astrometry. 

\texttt{orbitize!} is a Python package that streamlines the process of modeling the orbits of directly-imaged planets. The \texttt{orbitize!} development team aims to make orbit-fitting faster and more accessible by combining multiple existing algorithms and techniques into one code base \citep{Blunt_2020}. \texttt{orbitize!} currently models planet-planet interactions using a Keplerian approximation \citep{2021Lacour}, meaning it assumes planets' orbital elements stay the same over time. In this study, we document the addition of an N-body backend, REBOUND (\citealt{2012Rebound}, \citealt{reboundias15}) into \texttt{orbitize!}, which is now available in the 2.0.0 release. This update allows us to reliably model the motion of multiple secondary bodies, and we can use it to study when the current Keplerian approximation will be insufficient to model the motion of the HR 8799 system. Because it is a system with four super-Jupiter planets  \citep{2008Marois}, HR-8799 is an excellent candidate for directly observing the effects of planet-planet interactions. 

\section{Methods \& Results}
\label{sec:section2label}
\begin{figure*}
    \centerline{\includegraphics[width=0.75\linewidth]{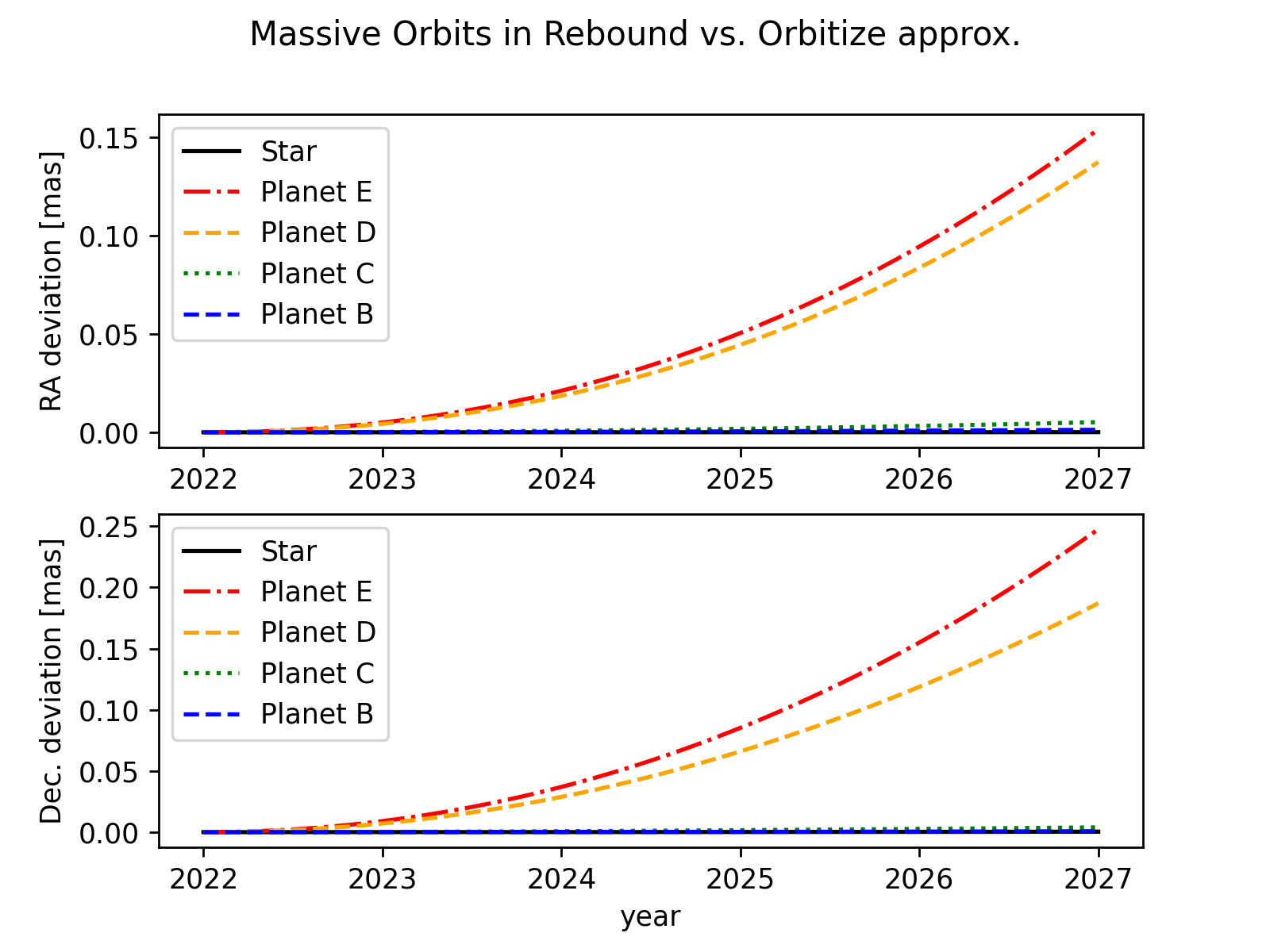}}
    \caption{The absolute difference between using an approximate Keplerian multi-body solver \citep{2021Lacour} vs. an N-body solver for each body in the HR-8799 system. Data shows that smaller-separation planets will deviate by up to a quarter of a milliarcsecond from the previous Kepler calculations within five years; enough to be detected by high-precision instruments such as VLTI-GRAVITY.}
    \label{fig:myfigure}
\end{figure*}

In order to confirm the accuracy of our REBOUND implementation, we first compared its output with the standard \texttt{orbitize!} Keplerian solver on a single massless secondary body. The massless results agreed to a factor of $10^{-10}$ fractional precision over hundreds of years, which is well below the measurement capabilities of current instruments. Since \texttt{orbitize!} and REBOUND use different orbital elements, our $10^{-10}$ agreement helps to confirm our conversions, and provides an independent check for our Kepler solver. After confirming that our N-body implementation was working, we moved on to testing massive systems against the current approximate solver in \texttt{orbitize!} (\autoref{fig:myfigure}), which shows deviations as high as 0.25 milliarcseconds for one body within 5 years (assuming the orbital parameters reported in \citealt{2018AJ}), detectable given the current VLTI-GRAVITY precision of 0.05 milliarcseconds \citep{Grav_coll_2019}. Within 10 years, the deviation is calculated to rise up to one milliarcsecond.

Although using an N-body solver has significant accuracy advantages for multi-planet systems, using it within \texttt{orbitize!} is completely optional, and can be called within the
\texttt{compute\_all\_orbits()} function. Anyone can use test data to experiment with the solvers, and tutorials are available on the \texttt{orbitize!} website\footnote{http://orbitize.info/}.

\subsection{Conclusion}
\label{sec:subsection2label}
Switching current Kepler or other mass-independent solvers for Newton-based N-body solvers such as REBOUND will be essential to model multi-planet orbits to the current level of precision of VLTI-GRAVITY. Within 5 years, we should be able to detect the differences between using a Kepler solver and an N-body solver in the HR-8799 system, which we will be able to use to make accurate dynamical mass measurements of directly-imaged exoplanets. With REBOUND's implementation into \texttt{orbitize!}, we can now look for planet-planet interactions in new and existing data, and use them to better predict astrometric motion.

\section*{Acknowledgements}
S.C., J.J.W., and S.B. acknowledge support from the Heising-Simons Foundation, including grant 2019-1698

Simulations in this paper made use of the REBOUND N-body code \citep{2012Rebound}. The simulations were integrated using IAS15, a 15th order Gauss-Radau integrator \citep{reboundias15}. 

\software{\texttt{orbitize!} v2.0.0 \citep{Blunt_2020}, \texttt{REBOUND} \citep{2012Rebound}, \texttt{astropy} (\citealt{astropy1}, \citealt{astropy2}), \texttt{numpy} \citep{numpy}, \texttt{matplotlib} \citep{matplotlib}.}


\bibliographystyle{aasjournal}
\bibliography{main}

\end{document}